\documentclass[prl,twocolumn]{revtex4-1}

\usepackage{amsmath, amssymb}
\usepackage{subfigure}
\usepackage{natbib}
\usepackage[colorlinks, 
linkcolor=blue,
urlcolor=blue,
citecolor=blue,
bookmarksopen=false]{hyperref}

\usepackage{amsmath}
\usepackage{graphicx, xcolor}
\usepackage{caption}

\newcommand{\bb}{\mathbf{b}}
\newcommand{\bB}{\mathbf{B}}

\newcommand{\bj}{\mathbf{j}}

\newcommand{\bu}{\mathbf{u}}
\newcommand{\bE}{\mathbf{E}}
\newcommand{\epol}{\mathbf{e}_\theta}

\newcommand{\fsa}[1]{\langle{#1}\rangle}
\newcommand{\jaa}[1]{#1}
\newcommand{\jab}[1]{#1}

\begin{document}

\title{Vanishing neoclassical viscosity and physics of the shear layer in stellarators}
\author{J.~L. Velasco, J.~A. Alonso, I. Calvo, {and J. Ar\'evalo}}
\affiliation{Laboratorio Nacional de Fusi\'on, {Asociaci\'on} EURATOM-CIEMAT, {28040} Madrid, Spain}
\email{joseluis.velasco@ciemat.es}
\homepage{\\ http://fusionsites.ciemat.es/jlvelasco}

\date{\today}

\begin{abstract}
  {The drift kinetic equation is solved for low density TJ-II plasmas
    employing slowly varying, time-dependent profiles. This allows to
    simulate density ramp-up experiments and describe from first
    principles the formation and physics of the radial electric field
    shear layer. The main features of the transition are perfectly
    captured by the calculation, and good quantitative agreement is
    also found. The results presented here, that should be valid for
    other non-quasisymmetric stellarators, provide a fundamental
    explanation for a wealth of experimental observations connected to
    the shear layer emergence in TJ-II. The key quantity is the
    neoclassical viscosity, which is shown to go smoothly to zero when
    the critical density is approached from below. This makes it
    possible for turbulence-related phenomena, and particularly zonal
    flows, to arise in the neighborhood of the transition.}
\end{abstract}

\maketitle

Understanding transport barriers in magnetic confinement
devices~\cite{burrell1997ExB} is a crucial issue for the fusion
programme {because they} allow access to regimes of improved
confinement in which commercial fusion reactors may be
viable. Confinement transitions have been observed and comprehensively
documented in a large variety of tokamaks and stellarators (see
Ref.~\cite{wagner2007hmode} and references therein). While the
reduction of turbulence by sheared flows is generally accepted as a
{key} ingredient in the transitions~\cite{terry2000shear}, the
identification of the {physical} mechanisms that create these sheared
flows is still an open issue. Zonal flows (i.e. turbulence-generated,
fluctuating and flux-surface collective radial electric fields) are
considered a possible catalyst of the transition, since they are
observed close before confinement transitions and they are able to
regulate transport~\cite{sugama2005prl}. In non-quasisymmetric
stellarators, the non-ambipolar radial fluxes, and consequently the
radial electric field, are basically determined by the neoclassical
theory of collisional transport in magnetized
plasma~\cite{helander2008amb}. Therefore, the interaction between
neoclassical and turbulent processes, and concretely the extent to
which turbulence can overcome the neoclassical viscosity and modify
the $E\times B$ rotation through momentum transport, has been the
subject of recent works
\cite{helander2008amb,sugama2005prl,itoh2007itb,alonso2012ishw}.  In
this Letter, we contribute to the above programme with the study of
the so-called low density confinement
transition~\cite{hidalgo2004pre,pedrosa2005eps,pedrosa2007sheared,pedrosa2008prl,Milligen11}
in the flexible heliac TJ-II~\cite{alejaldre1999first} which, as we
discuss below, shows similarities with other transitions described in
the literature. In this work the formation of the shear layer is
described from first principles in the framework of neoclassical
theory and the turbulent phenomena that arise in the neighborhood of
the transition are shown to be regulated by neoclassical transport.
TJ-II~\cite{alejaldre1999first} undergoes a spontaneous confinement
transition typically at a line-averaged electron density
$\overline{n_e}\!=\!n_{cr}\!\approx\!0.6\times 10^{19}\,$m$^{-3}$. At
this empirical critical density $n_{cr}$, the radial electric field
$E_r$ changes from positive to negative and, at the same time, a
transport barrier is generated close to the edge. The reversal of
$E_r$ starts where the density gradient is maximum
($\rho\!\approx\!0.8$, where $\rho\!=\!r/a$ is the normalized radius)
and then propagates {across} the entire region $0.5\!<\!\rho\!<\!0.9$
at a speed of the order of several m/s~\cite{happel2008twostep}. A
sheared $E_r$ always appears around the $E_r$-reversal point. Further
increase in $\overline{n_e}$ does not modify qualitatively $E_r$. {It
  is argued~\cite{pedrosa2007sheared} that the transport barrier is
  caused by turbulence reduction when the shearing rate of the
  $\bE\times\bB$ flow reaches values of the order of
  $10^{5}\,$s$^{-1}$; that is, when it becomes comparable to the
  estimated linear growth rate of the resistive interchange
  instability in TJ-II~\cite{carreras1993growthrate}.}

  Whereas similar transport barriers, related to jumps between roots
  of the ambipolar equation, have been observed in the low density
  regime of other stellarators~\cite{ida2008itb}, the very detailed
  study carried out in TJ-II has revealed additional interesting
  phenomena during the
  transition~\cite{pedrosa2008prl,pedrosa2005eps}. First of all, the
  level of turbulence and the $E \times B$ flux are seen to increase
  prior to the transition, and this was later shown to be associated
  to long-range-correlated (LRCed) electrostatic potential structures
  that grow when approaching the critical density. These experimental
  results have historically led to an interpretation of the transition
  in terms of the paradigm of sheared flow generation by Reynolds
  stress and turbulence reduction by shear, see
  e.g. Refs.~\cite{hidalgo2004pre,pedrosa2007sheared}. Along the same
  line of reasoning, the peaking of the potential relaxation time in
  biasing turn-off experiments~\cite{pedrosa2007sheared} and of the
  shear-flow susceptibility in electrode-biasing
  experiments~\cite{carralero2012suscep} were interpreted in terms of
  an increased turbulent drive during the transition. The increase of
  the level of turbulence and of the long-range correlations are
  neither specific of this transition nor of stellarators: they are
  also observed in the L-H transition of TJ-II (which happens at
  higher $\overline{n_e}$) and of other devices, see
  e.g. Ref.~\cite{xu2011lrc} and references therein. We note that
  there are important differences between the low-density and the L-H
  transition (e.g. the improvement of energy confinement in the
  latter). Nevertheless, the neoclassical modelling of the L-H
  transition~\cite{shaing1996prl} also relies on bifurcations of the
  ambipolarity condition.

  Since the inversion of $E_r$ is an essential part
  of the transition, a neoclassical study of the transition is in
  order. 
  The existence of multiple roots of the ambipolar
    equation is a well-known feature of stellarators in the
    low-density regime~\cite{shaing1984stab}. Indeed, a static neoclassical
    calculation of $E_r$ for plasmas immediately below (above)
    $n_{cr}$ was performed in Ref.~\cite{zurro2006rotation}, yielding
    positive (negative) $E_r$ in qualitative agreement with the
    experiment. In this work, we go beyond previous calculations and
    perform a full dynamical neoclassical calculation of the formation of the
    sheared $E_r$. We simulate the density ramp-up that leads to the
    transition and describe, from first principles,
      the formation and evolution of the shear flow in good agreement
    with the experiment. Furthermore, we show for the first time that
    the behaviour of the three quantities discussed
    above (amplitude of low-frequency LRCed potential
    fluctuations, potential relaxation time and shear-flow
    susceptibility)
        appear here as natural consequences of a neoclassical
        bifurcation. Namely, as the change of root approaches, the
      neoclassical poloidal viscosity (the restoring
        force of $E_r$ deviations towards its ambipolar value), goes
      to zero. {It will be shown that} this automatically implies a
      maximum of the low-frequency plasma potential
        fluctuations, of the relaxation times and of the
      shear-flow susceptibility.

  Let us begin by deriving the equation for the
  electric field evolution. We start from the momentum balance
  equation~\cite{helander2008amb} summed over species:
\begin{equation}
	m_i\frac{\partial (n\bu)}{\partial t} + \nabla\cdot\Pi_i + \nabla\cdot\Pi_e =  \bj\times\bB\,.~\label{EQ_MOMBAL}
\end{equation}
Here, $\bu$ is the ion flow tangent to flux surfaces, $\Pi_s$ is the
viscosity tensor, \jab{$\Pi_s = m_s\int
  \mathbf{v}\mathbf{v}f_s(\mathbf{x},\mathbf{v},t) d^3\mathbf{v}$}, or
momentum flux of species $s$, $f_s$ its distribution function, and
$\bj\times\bB$ is the Lorentz force. Other forces (such as neutral
friction) may be included as extra terms on the RHS of
Eq.~(\ref{EQ_MOMBAL}). We have assumed a {quasineutral plasma
  consisting of singly charged ions and electrons}
($n_e\!=\!n_i\!=\!n$). Note that we have dropped the inertia of the
electrons, given their much lower mass $m_s$, but kept the electron
viscosity tensor as it cannot be neglected in our low-$n$, high-$T_e$
plasmas~\cite{velasco2011er}. We work in Hamada magnetic coordinates
$(\psi, \theta, \xi)$, and follow the notation of
Ref.~\cite{alonso2012ishw}. The lowest order incompressible ion flow
is conveniently written as:
\begin{equation}\label{EQ_FLOW}
\bu = 2\pi\left(\frac{p_i'(\psi)}{ne} + \phi'(\psi)\right)\epol + \Lambda(\psi)\bB\,.
\end{equation}
The flux surface label $\psi$ is the toroidal magnetic flux and the prime
stands for derivative. The first term on the RHS of
Eq.~\eqref{EQ_FLOW} contains the diamagnetic and $E\times B$
perpendicular flows ($p_i$ is the ion pressure, $e$ the elementary
charge, $\phi$ is the electrostatic potential{, and} $\epol\times\bB =
(2\pi)^{-1}\nabla\psi$) together with the parallel Pfirsch-Schl\"uter
flow ($\nabla\cdot\epol = 0$ and $\fsa{\epol\cdot\bB} = 0$ for a
currentless stellarator). The term $\Lambda\bB$ is the ion bootstrap
flow~\cite{velasco2011bootstrap}. If we project Eq.~(\ref{EQ_MOMBAL})
{along} $\epol$ and take flux-surface-average $\fsa{\cdot}$ we obtain
our evolution equation for the radial electric field:
\begin{eqnarray}
 \frac{\partial E_r}{\partial t} &=& \frac{1}{n}\frac{\partial}{\partial t}\left(\frac{p_i'(r)}{e}\right)
  - E_r\frac{1}{n}\frac{\partial n}{\partial t} +
  \nonumber\\&+&\frac{(\psi'(r))^2}{4\pi^2 mn\fsa{\epol\cdot\epol}}\left(e(\Gamma_e-\Gamma_i)+\fsa{\bj\cdot\nabla r}\right)\,,\label{EQ_ER}
\end{eqnarray}
where $E_r\equiv -\phi'(r)$ and the minor radius $r$ is a geometric flux label
defined in terms of the volume $V(r)\!\equiv\!\pi r^2L_\textrm{ax}$, \jab{where $L_\textrm{ax}$ is the length of the magnetic axis}{. We} have obtained the
radial particle fluxes from
$\Gamma_s\!=\!-\frac{2\pi}{q_s\psi'(r)}\fsa{\epol\cdot\nabla\cdot\Pi_s}${~\cite{hirshman1981ff}.}
The viscosity tensor {can be split} into a neoclassical part, given by
the gyrotropic pressure tensor, and an anomalous contribution:
\begin{equation}
\Pi_s = \Pi_s^{NC} + \Pi_s^{an} = p_{s\|}\bb\bb + p_{s\perp}\left(\mathbf{I}-\bb\bb \right) + \Pi_s^{an}\,.\label{EQ_PI}
\end{equation}
As {mentioned above}, in non-quasisymmetric confining magnetic
topologies~\cite{helander2008amb}, the leading order contribution to
Eq.~(\ref{EQ_ER}) is $\fsa{\epol\cdot\nabla\cdot\Pi_s^{NC}}$, being
much larger than $\fsa{\epol\cdot\nabla\cdot\Pi_s^{an}}$, which will
be therefore neglected. Higher order neoclassical terms like the
shear-flow viscosity~\cite{helander2008amb} can be considered. This
term arises from orbit deviations away from flux surfaces and
transforms Eq.~(\ref{EQ_ER}) into a non-local diffusion equation for
$E_r$. Since this complicates the discussion and does not
fundamentally modify the predictions~\cite{turkin2011predictive}, we
neglect it and provide a posteriori justification based on the
experimental results.

\begin{figure}
\begin{center}
\includegraphics[angle=0,width=0.8\columnwidth]{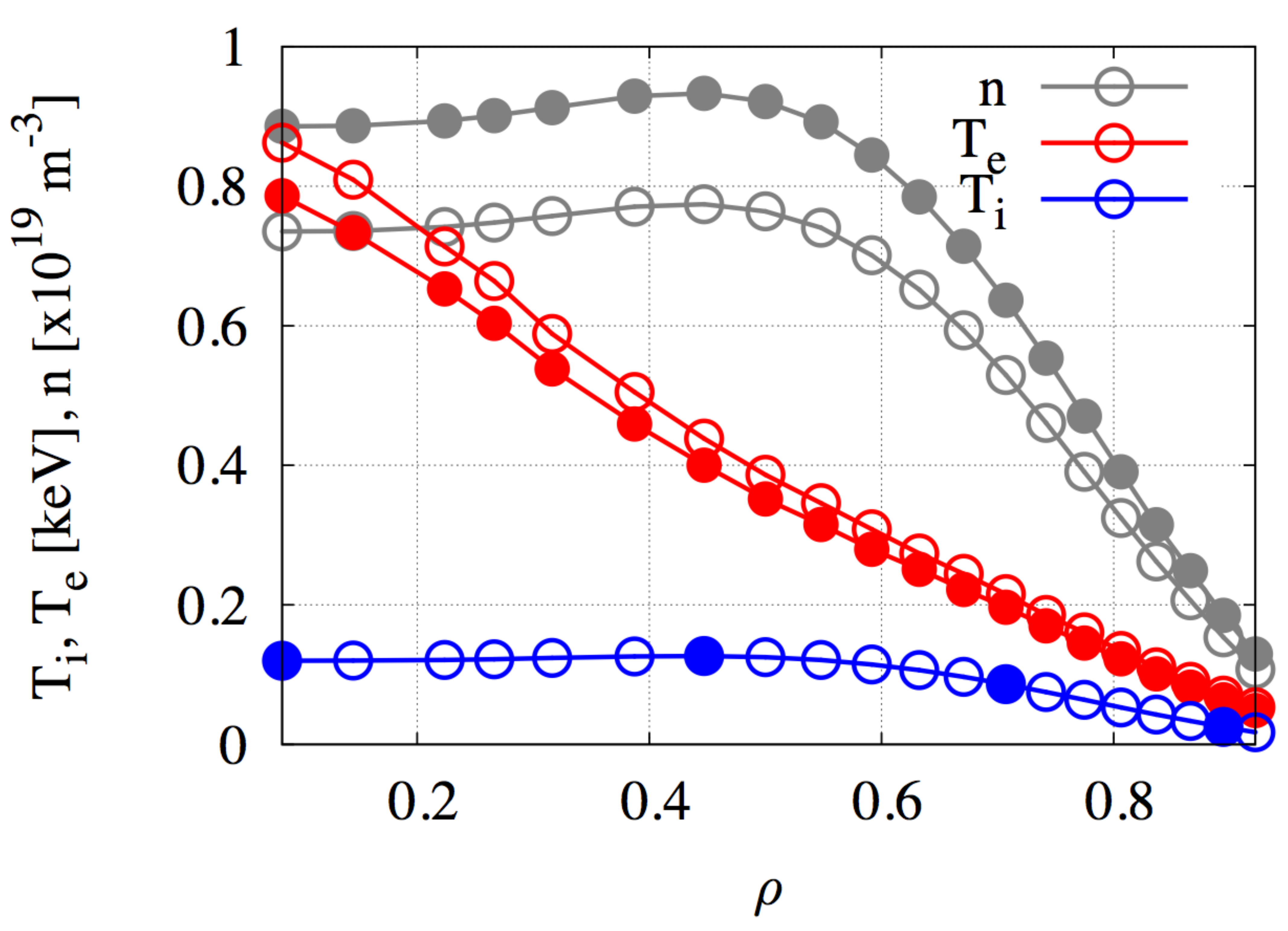}
\end{center}
\caption{(Color online) Plasma radial profiles for selected times: low (high) $n$
  in open (closed) circles. The evolution is given through $\frac{1}{n}\frac{\partial
    n}{\partial t}\!=\!3\,\mathrm{s}^{-1}\,$, $\frac{1}{T_e}\frac{\partial
    T_e}{\partial t}\!=\!- 1.5\,\mathrm{s}^{-1}\,$, and
  $\frac{\partial T_i}{\partial t}\!=\!0\,$.}
\label{FIG_PROF}
\end{figure}

We use the Drift Kinetic Equation Solver (DKES), complemented with
momentum-correction techniques, to evaluate the pressure anisotropy in
the TJ-II magnetic field in the parameter range usually found
{experimentally} in the vicinity of the transition. Details of the
calculation and convolution of the monoenergetic coefficients may be
found in Ref.~\cite{velasco2011bootstrap} and references therein. From
Eq.~\eqref{EQ_ER}, the time-evolution of $E_r$ is fully determined if
we know, at every instant of time, the magnetic configuration $b_{mn}$
and the profiles $n$, $T_e$, $T_i$.  Since we simulate a pure
proton-electron plasma~\cite{velasco2011er}, the effective charge
$Z_{eff}$ (which mainly affects collisionality) is set equal to
one. We perform a numerical simulation of a density ramp {across} the
critical density for a plasma with profiles $n(\rho,t)$, $T_e(\rho,t)$
and $T_i(\rho,t)$ that mimic the experimental {ones}, see
Fig.~\ref{FIG_PROF} and Refs.~\cite{pedrosa2005eps,zurro2006rotation}.
We set $\fsa{\bj\cdot\nabla r}\!=\!0$ unless otherwise stated. This is
implied by quasineutrality ($\nabla\cdot\bj = 0$), but a net radial
plasma current can be induced in plasma biasing experiments. An
important point to be noted is that, although we assume that the
leading non-ambipolar particle fluxes are neoclassical, we make no
particular assumption on the total particle or energy fluxes. They are
included (together with the sources) implicitly in the evolution of
$n$, $T_e$, $T_i$.

\begin{figure}
\begin{center}
\includegraphics[angle=0,width=0.8\columnwidth]{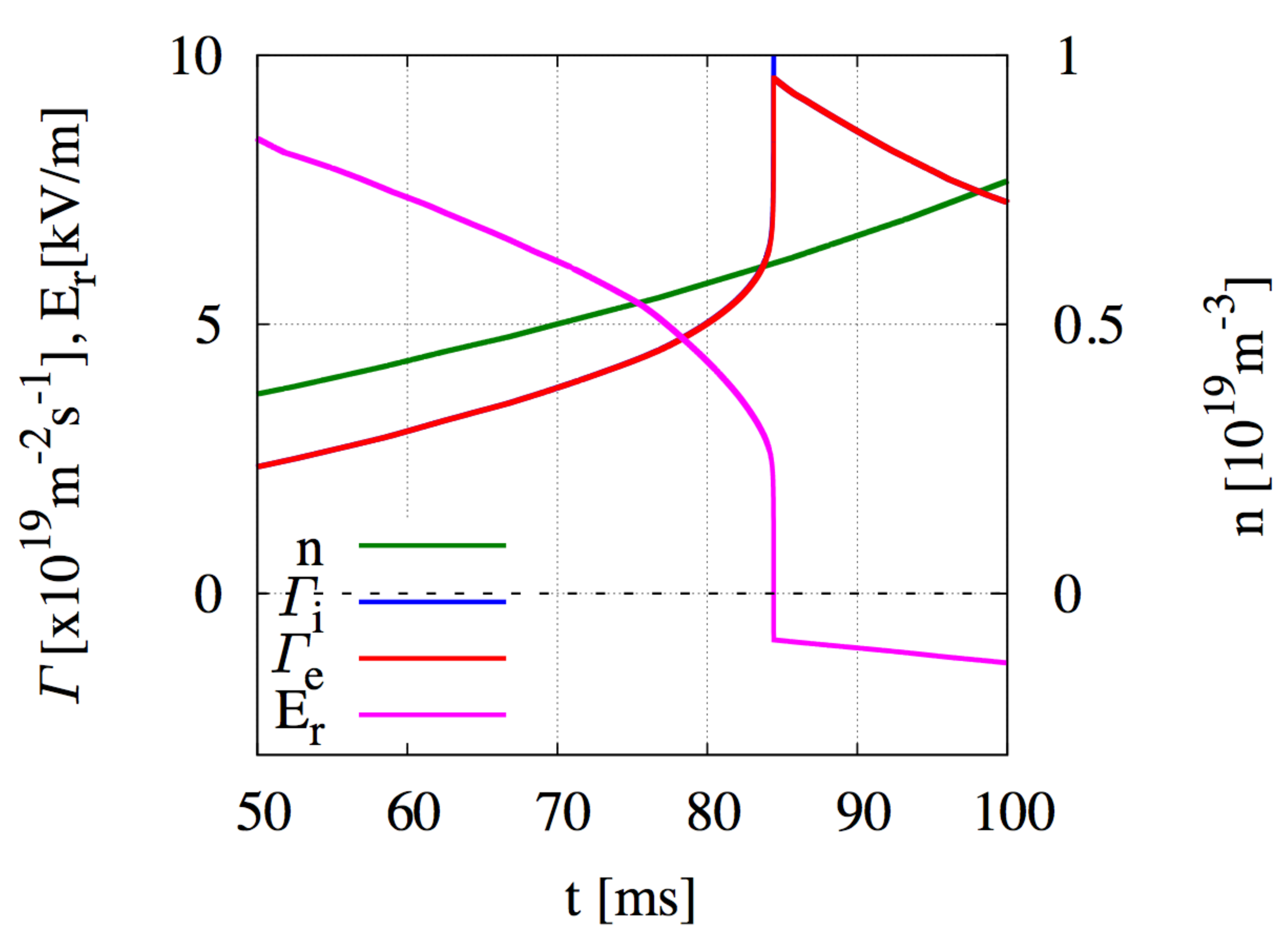}
\end{center}
\caption{(Color online) Time evolution of the relevant quantities at $\rho\!=\!0.7$ during the density ramp-up.}
\label{FIG_EVOL}
\end{figure}

\begin{figure}
\begin{center}
\includegraphics[angle=0,width=0.8\columnwidth]{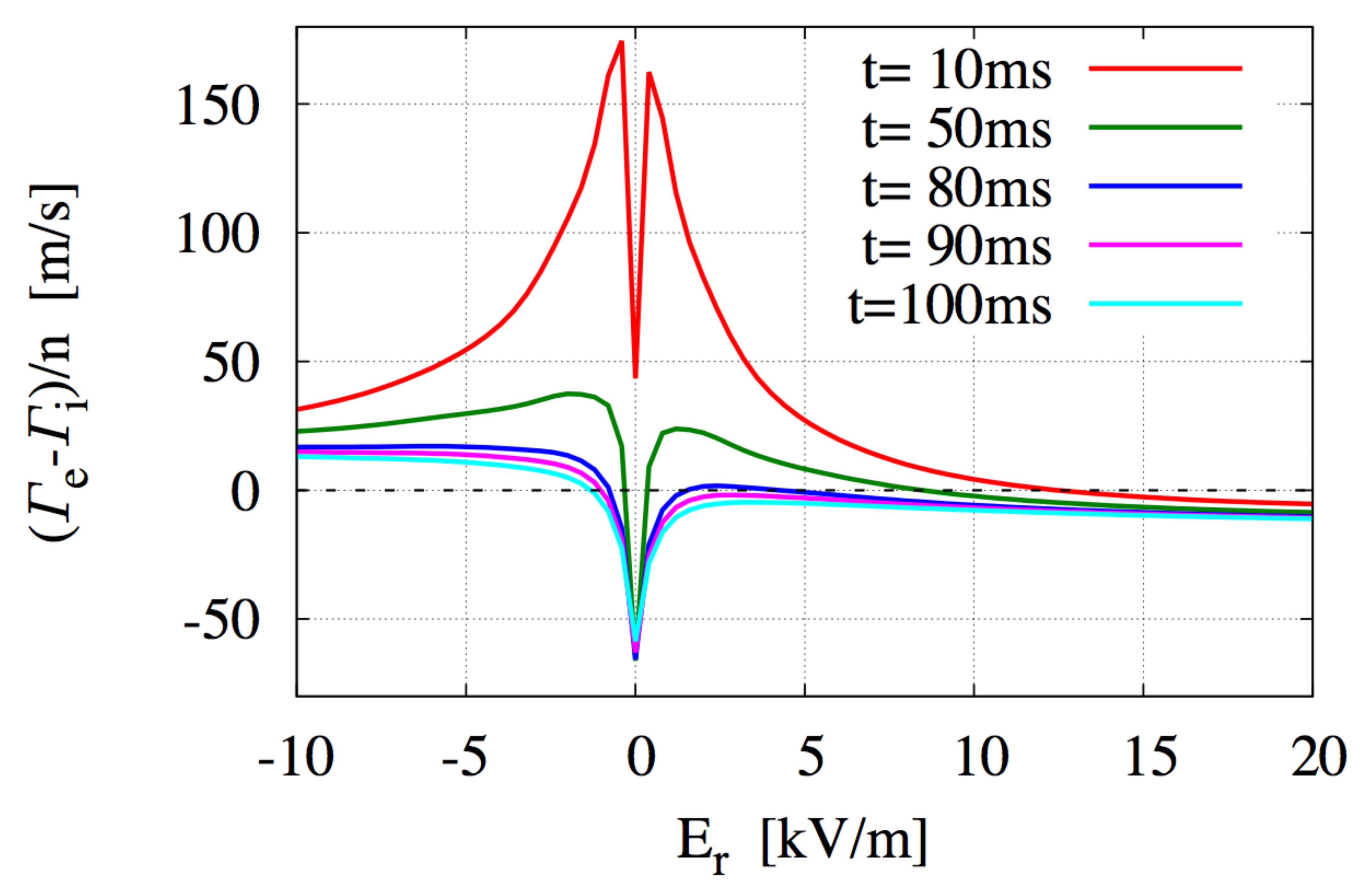}
\end{center}
\caption{(Color online) Ambipolar equation at $\rho\!=\!0.7$ for several
  representative times.}
\label{FIG_AMB}
\end{figure}

\begin{figure}
\begin{center}
\includegraphics[angle=0,width=0.8\columnwidth]{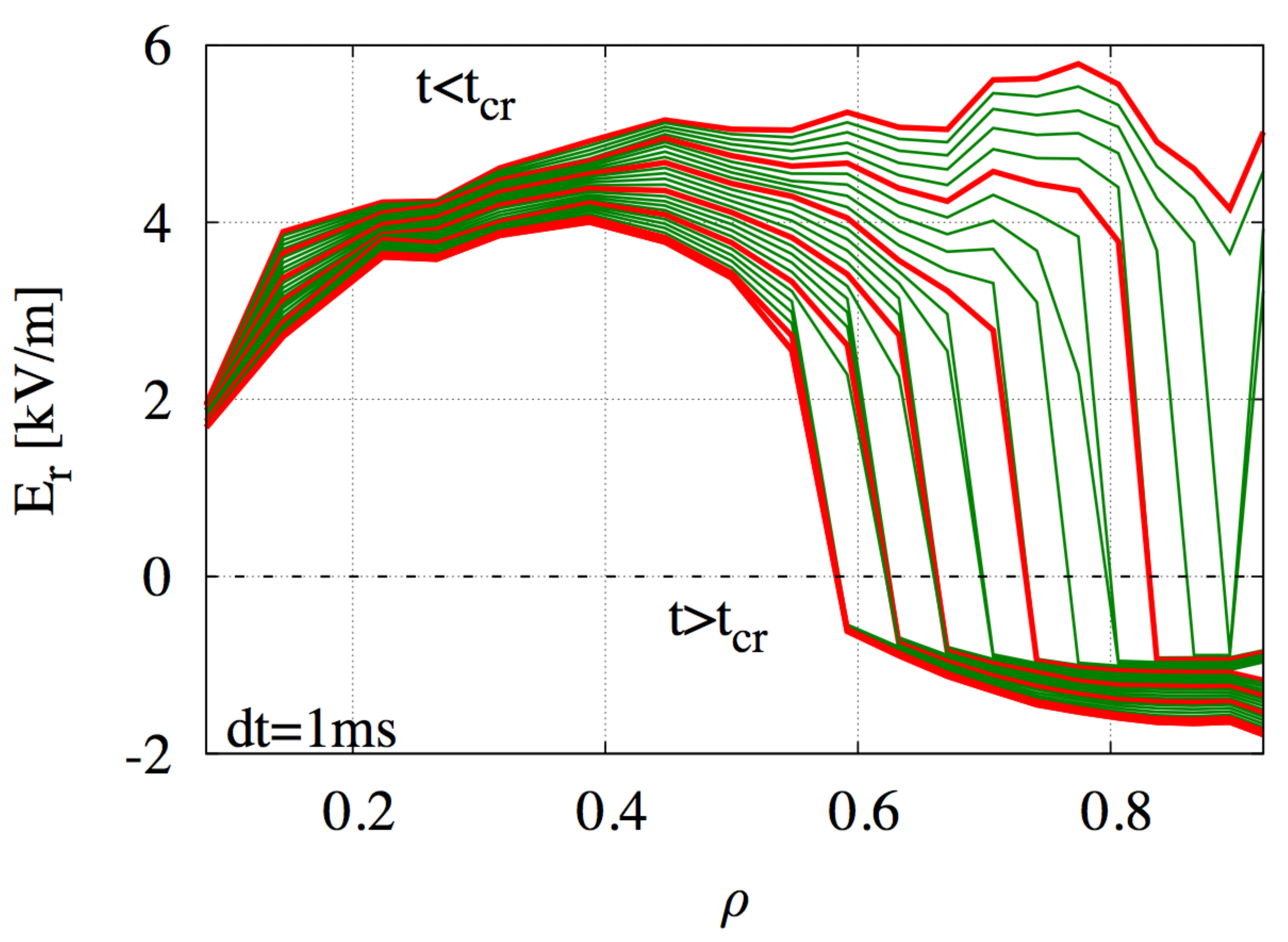}
\end{center}
\caption{(Color online) Radial electric field profile for representative times. Lines
  are separated by 1$\,$ms and thick lines by 5$\,$ms. The starting
  ($t\!<\!t_{cr}$) $E_r(\rho)$ is positive.}
\label{FIG_ER}
\end{figure}

In Fig.~\ref{FIG_EVOL} we sketch the evolution of the main quantities
of our numerical ramp-up experiment in a representative radial
position, $\rho\!=\!0.7$. The evolution of $E_r$ is given only by the
evolution of $n$ and $T_e$ during the ramp-up. As $n$ rises, the
radial fluxes become larger and $E_r$ less positive. $\Gamma_i$ is
slightly higher than $\Gamma_e$, which sets the sign of $\partial
E_r/\partial t$, but the difference is not even visible in the
figure. This is a consequence of the very slow variation of $n$ and
$T_e$, and will allow us to discuss some of the
results in terms of the steady-state ambipolar equation. For
$n\!\approx\!0.6\times 10^{19}\,$m$^{-3}\,$, there is a change of
root: $E_r$ goes from positive to negative in several tens of
$\mu$s. For larger $n$, further increase in $n$ leads to
smaller $\Gamma_s$ and more negative
$E_r$. This change of behaviour of $\Gamma_s$ (they
  grow with $n$ below $n_{cr}$ and decrease above $n_{cr}$) might
  indicate a neoclassical particle transport barrier (nevertheless, let us
  remember that it is probably the ambipolar anomalous
  flux~\cite{pedrosa2005eps} which controls particle transport).

  This general behaviour is expected for neoclassical simulations of
  TJ-II low-density plasmas~\cite{velasco2011er,zurro2006rotation}. In
  Fig.~\ref{FIG_AMB}, we show the ambipolar equation at $\rho\!=\!0.7$
  for several relevant times.  Since we start from low
  {collisionality,} $E_r$ is positive ($t\!=\!10\,$ms). As $n$ is
  raised, a negative stable root appears ($t\!=\!50\,$ms and
  $t\!=\!80\,$ms), but the two stable roots are separated by an
  unstable root, so $E_r$ stays positive. Only for $n\!=\!n_{cr}$,
  when the electron root disappears ($t\!=\!90\,$ms and
  $t\!=\!100\,$ms), the system {\em jumps} to negative $E_r$ in tens
  of $\mu$s (the typical time of evolution towards ambipolarity, see
  below). Fig.~\ref{FIG_ER} shows, for the first time, the formation
  and precise evolution of the shear layer in TJ-II plasmas. It starts
  to develop at $\rho\!=\!0.85$, approximately where the gradient is
  maximum~\cite{pedrosa2005eps} and then propagates inwards and
  outwards: a speed of the order of 1$\,$m/s may be extracted, as
  measured in Ref.~\cite{happel2008twostep}. Let us recall that, since
  we neglect the shear-flow viscosity, our simulations are local,
  hence the evolution equation for $E_r$ is solved independently at
  each radial position, and only an indirect coupling exists through
  the $n$ and $T_e$ gradients. Thus, the speed of the layer
  propagation is determined basically by the evolution time of the
  local collisionality (as we have discussed, $n$ and $T_e$). Shearing
  rates of the order of $10^5\,$s$^{-1}$ can be inferred from
  Fig.~\ref{FIG_ER}, so this neoclassical shear might be large enough
  for playing a role in the reduction of turbulent transport (the
  linear growth rate of the resistive interchange instability, that is
  thought to be the dominant one at the edge of TJ-II, is also of the
  order of $10^5\,$s$^{-1}$~\cite{carreras1993growthrate}.

  \jaa{ The main features of the low density transition are well
    captured by neoclassical transport and the correct prediction of
    the shear layer supports the initial ordering assumption. If the
    shear-flow viscosity had been calculated, the shear of $E_r$ would
    be somewhat smaller~\cite{turkin2011predictive}, but this
    agreement with the experiment shows that no qualitative
    differences would have been found.}

    Generally, this picture is only slightly modified by considering
    the measured turbulent momentum transport. Indeed, local
    measurements of Reynolds stress~\cite{alonso2012ishw} show average
    events $\Gamma_{RS}/n\!\lesssim\!0.1\,$m/s (which would not be
    perceptible in the evolution of $E_r$, see Fig.~\ref{FIG_AMB}) and
    infrequent extreme events of $\Gamma_{RS}/n\!\lesssim\!10\,$m/s
    (which would not fundamentally modify the results). This is
    however not true when the transition is approached from below as
    can be seen in Fig.~\ref{FIG_AMB}. In this situation ($t =
    80$~ms), the non ambipolar neoclassical fluxes display a weak
    dependence on $E_r$ around the ambipolar value and large $E_r$
    excursions may be caused by turbulent momentum fluxes or external
    forcing (biasing) as is observed experimentally.  
\begin{figure}
\begin{center}
\includegraphics[angle=0,width=0.8\columnwidth]{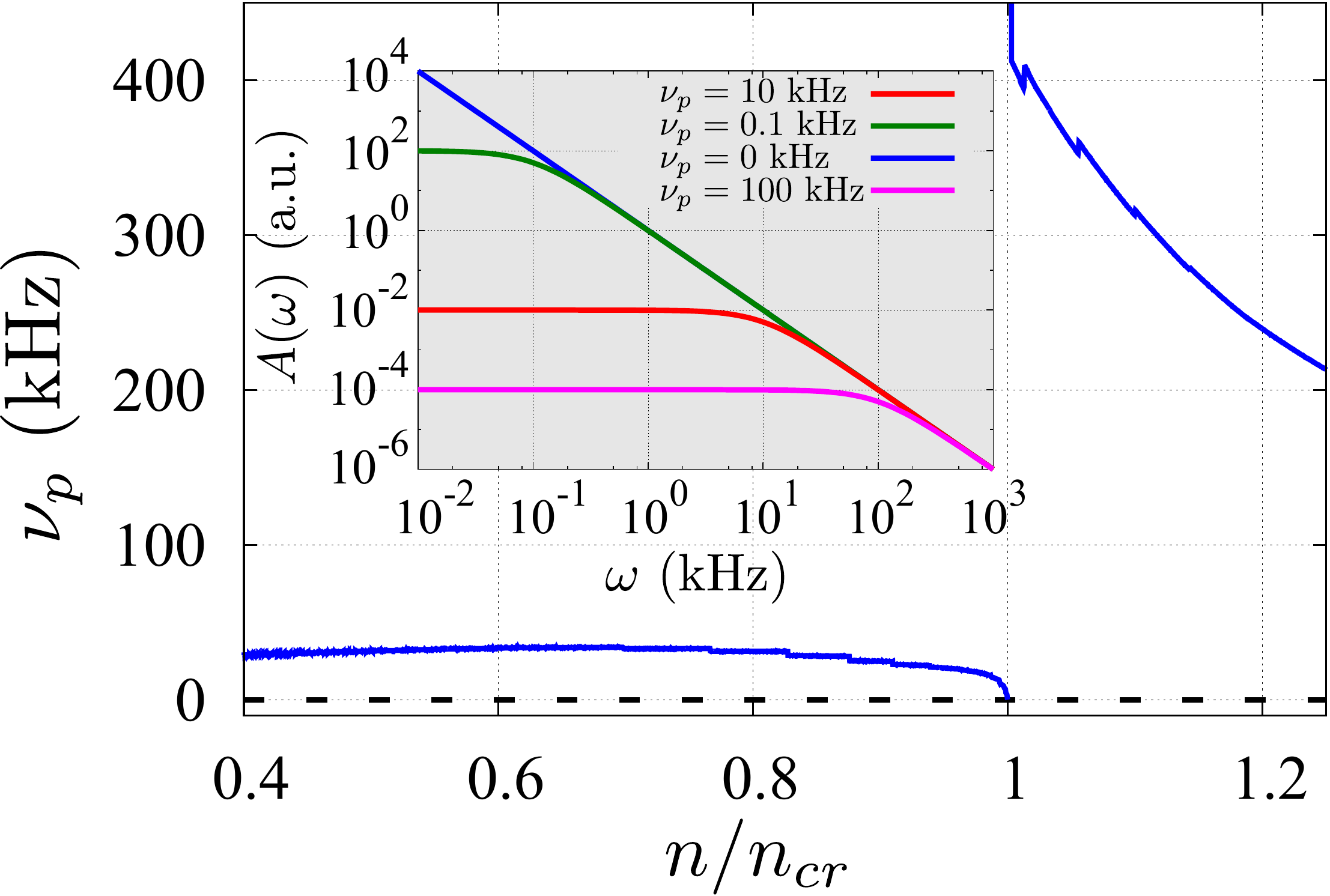}
\end{center}
\caption{(Color online) Density dependence of the neoclassical poloidal viscosity during the
  transition. $n_{cr}$ is locally defined as the $n$ at which $E_r$
  passes through 0 (inset: frequency dependence of the neoclassical damping for
  several values of $\mu_p$).}
\label{FIG_VISCO}
\end{figure}

\jaa{
To make the argument more precise we define a neoclassical poloidal viscosity as the linear coefficient of the difference between the
electron and ion radial fluxes (shown in Fig.~\ref{FIG_AMB}) expanded around the ambipolar
electric field,} $[\Gamma_e-\Gamma_i](E_r)\!=
-\mu_p(E_r-E_r^0)+O((E_r-E_r^0)^2)$. Eq.~(\ref{EQ_ER}) then yields:
\begin{eqnarray}
\frac{\partial E_r}{\partial t}  &&\approx 
  \frac{e(\psi')^2}{4\pi^2 mn\fsa{\epol\cdot\epol}}
\Bigg[\mu_p(E_r-E_r^0) -\frac{\fsa{\bj\cdot\nabla r}}{e}\Bigg] \nonumber\\&&=-\nu_p(E_r-E_r^0)
+\check{j_r}\,.\label{EQ_NUP}
\end{eqnarray}
For the sake of argument we have neglected the
  (assumed slow) {\jab{variations in $n$ and $T_i$. The coefficient}
    $\mu_p$ can be calculated directly from the data of
    Fig.~\ref{FIG_EVOL}, and we have {absorbed} $n$, $m$, $e$,
    $\psi'${, $\fsa{\epol\cdot\epol}$, and} constants into $\nu_p$ and
    $\check{j_r}$. {This} scalar viscosity is
    a combination of the elements in the viscosity matrix defined in
    Ref.~\cite{sugama2002viscosity} (in particular, in the limit where
    the parallel currents are low and the temperature gradients small,
    it is proportional to the $L_1$ coefficient). The dependence of
    $\nu_p$ on $n$ during the transition is shown in
    Fig.~\ref{FIG_VISCO}: it is \jaa{smaller before the transition
      than after it} (as predicted, with a simplified formulation, in
    Ref.~\cite{itoh2007itb}) and, more importantly, goes to zero when
    approaching $n_{cr}$ from $n\!<\!n_{cr}$. \jaa{We now show that
      the behaviour of the neoclassical viscosity provides a simple,
      unified explanation of the observed} phenomena (amplitude of
    low-frequency, LRCed potential fluctuations, potential relaxation
    time and shear-flow susceptibility)} that accompany the
  transition. To our knowledge, this is the first time that the
  vanishing of the neoclassical viscosity is explicitly shown to be
  enough to cause the phenomena {\em at} the transition.

\jaa{
The characteristic relaxation time in biasing turn-off experiments ($\nu_p^{-1}$ in Eq.~\ref{EQ_NUP}) shows a peak around the critical density and decreases for larger density plasmas in the ion root \cite{pedrosa2007sheared}. This is reproduced by the curve shown in Fig.~\ref{FIG_VISCO}. Similarly, when a low frequency external biasing is applied, the response electric field is in phase with the biasing and its amplitude increases close to the critical density \cite{carralero2012suscep}. This is to be expected from Eq.~\ref{EQ_NUP}, for in that case $E_r(t) \approx E_r^0+\nu_p^{-1}\check{j_r}(t)$. 
Finally, to better discuss the observations of LRCs close to the transition~\cite{hidalgo2004pre} we Fourier
  transform Eq.~(\ref{EQ_NUP})}:
\begin{eqnarray}
&&\hspace{-0.5cm}  i\omega\hat{E_r}(\omega) = -\nu_p\hat{E_r}(\omega) +\hat{j}(\omega)
\Rightarrow\nonumber \\
&&  |\hat{E_r}(\omega)|^2 = \frac{1}{\nu_p^2+\omega^2}|\hat{j}(\omega)|^2 \equiv
  A(\omega)|\hat{j}(\omega)|^2\,,\label{EQ_OMEGA}
\end{eqnarray}
for time scales faster than that of the density ramp, i.e.,
$\omega\!>\!\partial_t\log(E_r^0)\!\sim\!\partial_t\log(n)\!\sim\!10\,$Hz.
Eq.~\ref{EQ_OMEGA} shows that the amplitude of the fluctuations
$\hat{E_r}(\omega)$ driven by a given \jaa{broadband turbulent forcing
  $\hat{j}(\omega)$ is modulated by the neoclassical viscosity, which
  damps fluctuations of frequencies lower than $\nu_p$.} Below and
above the transition (Fig.~\ref{FIG_VISCO} inset), the fluctuations
$\hat{E_r}(\omega)$ with $\omega\!\lesssim\!10\,$kHz are
neoclassically damped. It is only close below $n_{cr}$ that the
neoclassical viscosity drops, leaving the low frequency $E_r$
fluctuations \jaa{(which are expected to display higher LRC)}
undamped.  \jab{A higher order shear-flow
  viscosity~\cite{turkin2011predictive} is usually included in
  Eq.~(\ref{EQ_ER}), whose main effect is to smooth the radial
  variations of the $E_r$. For the present discussion, this term might
  only be important in the inner side of the shear layer, where the
  leading order viscosity goes to zero, and will set a low (though not
  null) effective viscosity and damping rate. } The experimental
measurement~\cite{pedrosa2007sheared} of a peaking factor of about 5
in $\nu_p^{-1}$ gives an upper limit to these second order effects.

  Note that the vanishing of the viscosity is a consequence of the
  transition from electron to ion root and the above phenomena are not
  to be expected in the reverse transition~\cite{ida2008itb,
    Milligen11}.

In conclusion, we have described, by solving the drift kinetic
equation, the formation of the shear layer in TJ-II. Even though such
a first-principles theoretical calculation is by itself an important
result, we have additionally advanced in the understanding of relevant
physical phenomena associated to the transition, and also observed in
other devices. They are essentially related to the
amplification of dynamical electric fields as a result of the
reduction or vanishing of the neoclassical viscosity.

Discussions with C. Hidalgo and the collaboration of the TJ-II
team are acknowledged. This research was supported in part by grant
ENE2009-07247, Ministerio de Ciencia e Innovaci\'on (Spain).

\end{document}